\documentclass[prd,twocolumn,amsmath,amssymb,axodraw]{revtex4}
\usepackage{graphicx}
\begin{document}

\title{\boldmath Search for $D^+ \to \phi l^+ \nu$
and measurement of the branching fraction for $D^+ \to \phi \pi^+$}

\author{
M.~Ablikim$^{1}$,      J.~Z.~Bai$^{1}$,            Y.~Ban$^{11}$,
J.~G.~Bian$^{1}$,      X.~Cai$^{1}$,               H.~F.~Chen$^{15}$,
H.~S.~Chen$^{1}$,      H.~X.~Chen$^{1}$,           J.~C.~Chen$^{1}$,
Jin~Chen$^{1}$
% ...
\\
\vspace{0.5cm}
(BES Collaboration)\\
\vspace{0.5cm}
{\it
$^{1}$ Institute of High Energy Physics, Beijing 100049, People's Republic
of China\\
% ...
$^{11}$ Peking University, Beijing 100871, People's Republic of China\\
% ...
$^{15}$ University of Science and Technology of China, Hefei 230026,
People's Republic of China\\
% ...
}}


\vspace{1.0cm}
\begin{abstract}

Using a data sample of integrated luminosity of about 33 pb$^{-1}$
collected around 3.773 GeV with the BESII detector at the BEPC collider,
the semileptonic decays $D^+ \to \phi e ^+\nu_e$,
$D^+ \to \phi \mu^+\nu_\mu$ and the hadronic decay
$D^+ \to \phi \pi^+$ are studied. The upper limits of the branching
fractions are set to be $BF(D^+ \to \phi e ^+\nu_e) <$ 2.01\% and
$BF(D^+ \to \phi \mu^+ \nu_\mu) <$ 2.04\% at the 90\% confidence level.
The ratio of the branching fractions for $D^+ \to \phi \pi^+$
relative to $D^+ \to K^-\pi^+\pi^+$ is
measured to be $0.057 \pm 0.011 \pm 0.003$. In addition,
the branching fraction for $D^+ \to \phi \pi^+$ is obtained
to be $(5.2 \pm 1.0 \pm 0.4) \times 10^{-3}$.
\end{abstract}

\maketitle

\section{Introduction}

Searching for new modes in charm decays is of great
interest. It not only investigates possible decay mechanism
and finds its contribution to the total decay width, but
is also useful to simulate accurately cascade decays of bottom mesons
and to eliminate backgrounds of charm modes in studying
bottom decays.

In this paper, the semileptonic and hadronic $D$ decays in which
the final state particles contain a $\phi$ meson are studied.
Whenever a specific state or decay
mode is mentioned in this work, the charge-conjugate state or
decay mode is always implied.

\section{BESII Detector}

The BESII detector upgraded from the BES~\cite{bes}
is a large solid-angle magnetic spectrometer described in detail
elsewhere~\cite{besII}. A 12-layer Vertex Chamber (VC) surrounding
the beryllium beam pipe provides the trigger and
coordinate informations. A forty-layer main drift chamber (MDC),
located outside the VC, yields precise measurements of charged
particle trajectories with a solid angle coverage of $85\%$ of
$4\pi$; it also provides ionization energy loss ($dE/dx$)
measurements used for particle identification. Momentum resolution
of $1.78\%\sqrt{1+p^2}$ ($p$ in GeV/c) and $dE/dx$
resolution of $8.5\%$ for Bhabha scattering are
obtained for data taken at $\sqrt s$ = 3.773 GeV.
An array of 48 scintillation counters surrounding the MDC measures
the time of flight (TOF) of charged particles with a resolution of
about 180 ps for electrons. Outside the TOF is a 12 radiation length
barrel shower counter (BSC) comprised of gas tubes
interleaved with lead sheets. The BSC measures the energies
of electrons and photons over $80\%$ of the total solid angle
with an energy resolution of $\sigma_E/E = 0.22/\sqrt E$
($E$ in GeV) and spatial resolution of $\sigma_\phi = 7.9$ mrad
and $\sigma_Z = 2.3$ cm for electrons. A solenoidal
magnet outside the BSC provides a 0.4 T magnetic
field in the central tracking region of the detector.
The magnet flux return is instrumented with
three double layers of counters, that are used to
identify muons with momentum greater than 500 MeV/c and
cover $68\%$ of the total solid angle.

\section{Data analysis}

The data used for this analysis were collected
around the center-of-mass energy of 3.773 GeV
with the BESII detector operated at the Beijing Electron Positron Collider
(BEPC). The total integrated luminosity of the data set is about 33
pb$^{-1}$. At the center-of-mass energy 3.773 GeV, the
$\psi(3770)$ resonance is produced in electron-positron
($e^+e^-$) annihilation. The $\psi(3770)$ decays predominately
into $D\bar D$ pairs. If one $\bar D$ meson is fully reconstructed,
the $D$ meson must exist in the system recoiling against
the fully reconstructed $\bar D$ meson (called singly tagged $\bar D$).
Using the singly tagged $D^-$ sample, the semileptonic decays
$D^+ \to \phi e^+ \nu_e$ and $D^+ \to \phi \mu^+ \nu_\mu$ are searched
in the recoiling system. The hadronic candidates $D^+ \to \phi \pi^+$
and $D^+ \to K^- \pi^+\pi^+$
are reconstructed directly from the data sample of 33 pb$^{-1}$.

\subsection{Event selection}

Events which contain at least three charged tracks
with good helix fits are selected. To ensure good
momentum resolution and reliable charged particle identification,
every charged track is required to
satisfy $|$cos$\theta |<0.85$, where $\theta $ is the
polar angle.
All tracks, save those from $K^0_S$ decays, must originate from the
interaction region by requiring that the closest approach of a charged
track is less than 2.0~cm in the $xy$ plane and 20~cm in the $z$ direction.
Pions and kaons are identified by means of the
combined particle confidence level which is calculated
with information from the $dE/dx$ and TOF measurements~\cite{psipp1,psipp2}.
Pion identification requires a consistency with the pion hypothesis
at a confidence level ($CL_{\pi}$) greater than $0.1\%$.
In order to reduce misidentification, a kaon candidate is required
to have a larger confidence level ($CL_{K}$) for a kaon hypothesis
than that for a pion hypothesis.
For electron or muon identification, the combined particle confidence
level ($CL_e$ or $CL_\mu$), calculated for the $e$ or $\mu$ hypothesis
using the $dE/dx$, TOF and BSC measurements, is required to
be greater than $0.1\%$.

The $\pi^0$ is reconstructed in the decay of
$\pi^0 \to \gamma \gamma$. To select good photons from the $\pi^0$ decay,
the energy deposited in the BSC is required to be greater than 0.07 GeV,
and the electromagnetic shower is required to start in the first
5 readout layers. In order to reduce backgrounds, the angle between
the photon and the nearest charged track is required to be greater
than $22^\circ$, and the angle between the cluster development direction
and the photon emission direction to be less than
$37^\circ$~\cite{psipp1,psipp2}.

\subsection{Singly tagged $D^-$ sample}

\begin{figure}[htbp]
\begin{center}
\includegraphics[width=9cm,height=7cm]
{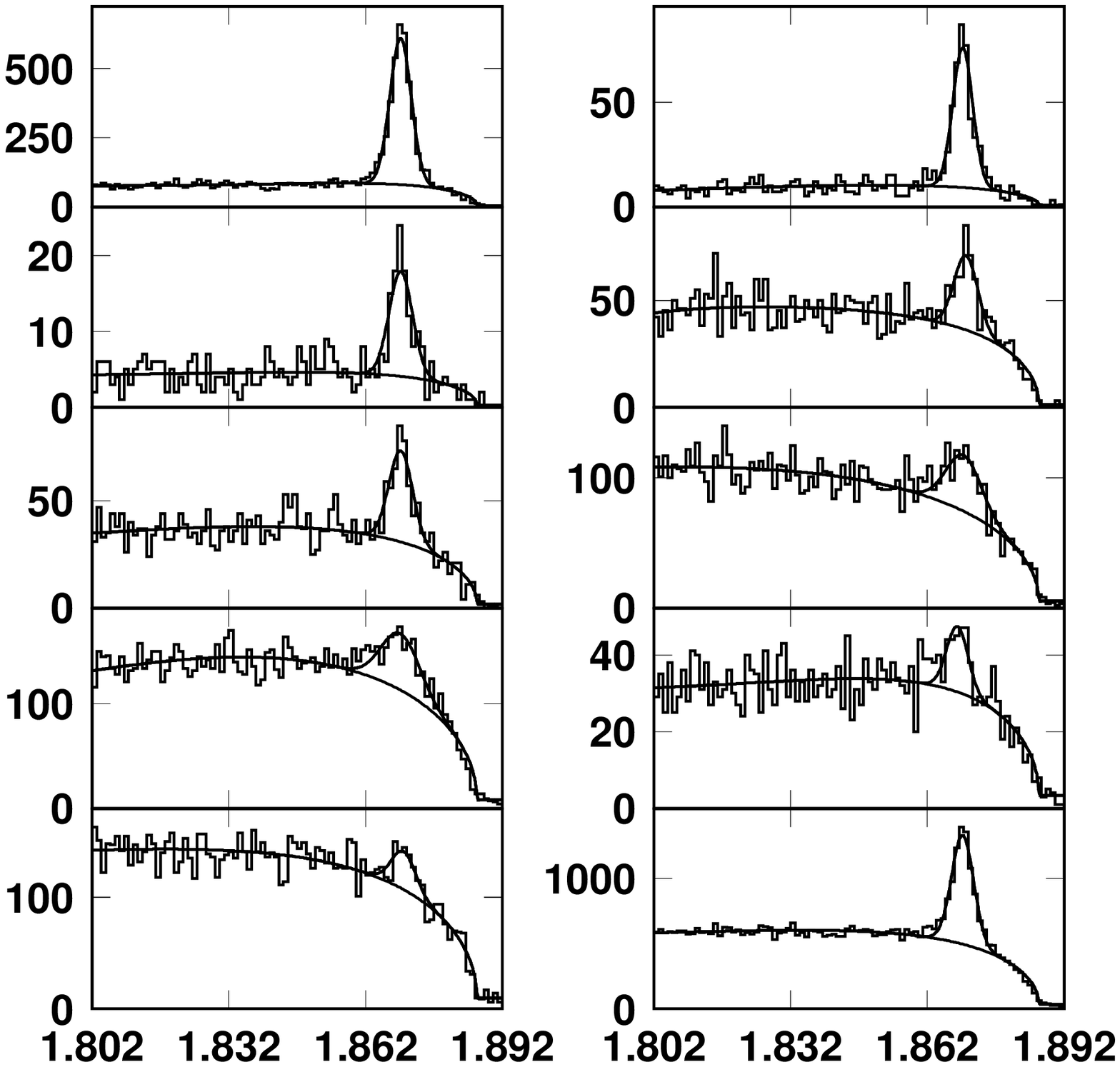}
\put(-210,175){\bf (a)}
\put(-90,175){\bf (b)}
\put(-210,145){\bf (c)}
\put(-90,145){\bf (d)}
\put(-210,115){\bf (e)}
\put(-90,100){\bf (f)}
\put(-210,65){\bf (g)}
\put(-90,62){\bf (h)}
\put(-210,35){\bf (i)}
\put(-90,45){\bf (j)}
\put(-150,0){Invariant Mass (GeV/$c^2)$}
\put(-250,50){\rotatebox{90}{Events/(0.001 GeV/$c^2$)}}
\caption{\small The distributions of the fitted invariant masses of
   (a) $K^+\pi^-\pi^-$, (b) $K^0_S \pi^-$, (c) $K^0_S K^-$,
   (d) $K^+ K^-\pi^-$,  (e) $K^0_S \pi^-\pi^-\pi^+$, (f) $K^0_S \pi^-\pi^0$,
   (g) $K^+ \pi^-\pi^-\pi^0$, (h) $K^+\pi^+\pi^-\pi^-\pi^-$,
   (i) $\pi^-\pi^-\pi^+$ combinations; (j) is the fitted masses of the
$mKn\pi$ combinations for the nine modes combined together.}
    \label{dptags_9modes}
\end{center}
\end{figure}

The singly tagged $D^-$ sample used in the analysis
was selected previously~\cite{psipp1,psipp2,psipp3,psipp4}.
The singly tagged $D^-$ mesons were reconstructed in
the nine hadronic modes of $D^- \to K^+\pi^-\pi^-$,
$K^0_S\pi^-$, $K^0_S K^-$, $K^-K^+\pi^-$, $K^0_S\pi^-\pi^-\pi^+$,
$K^0_S\pi^-\pi^0$, $K^+\pi^-\pi^-\pi^0$, $K^+\pi^-\pi^-\pi^-\pi^+$
and $\pi^-\pi^-\pi^+$. The distributions of the fitted invariant masses of
the $mKn\pi (m = 0, 1, 2; n = 1, 2, 3, 4)$ combinations are shown in
Fig.~\ref{dptags_9modes}.
The number of the singly tagged $D^-$ mesons is $5321 \pm 149 \pm
160$~\cite{psipp3,psipp4},
where the first error is statistical and the second systematic.

\subsection{Candidates for $D^+ \to \phi e^+\nu_e$ and $D^+\to \phi
\mu^+\nu_\mu$}

Candidates for $D^+\to \phi e^+\nu_e$ and
$D^+\to \phi \mu^+\nu_\mu$ are
selected from the surviving tracks in the system recoiling against
the tagged $D^-$. To select these candidates, it is required that there
are three charged tracks, one of which is identified as an electron or a
muon with the charge opposite to the charge of the tagged $D^-$ meson,
the other two are identified as a $K^+$ and a $K^-$ mesons. The $\phi$ meson
is selected by requiring that the difference between the invariant
mass of $K^+K^-$ and the nominal $\phi$ mass should be less than
20 MeV/$c^2$.

\begin{figure}[hbtp]
\begin{center}
\includegraphics[width=8cm,height=6cm]
{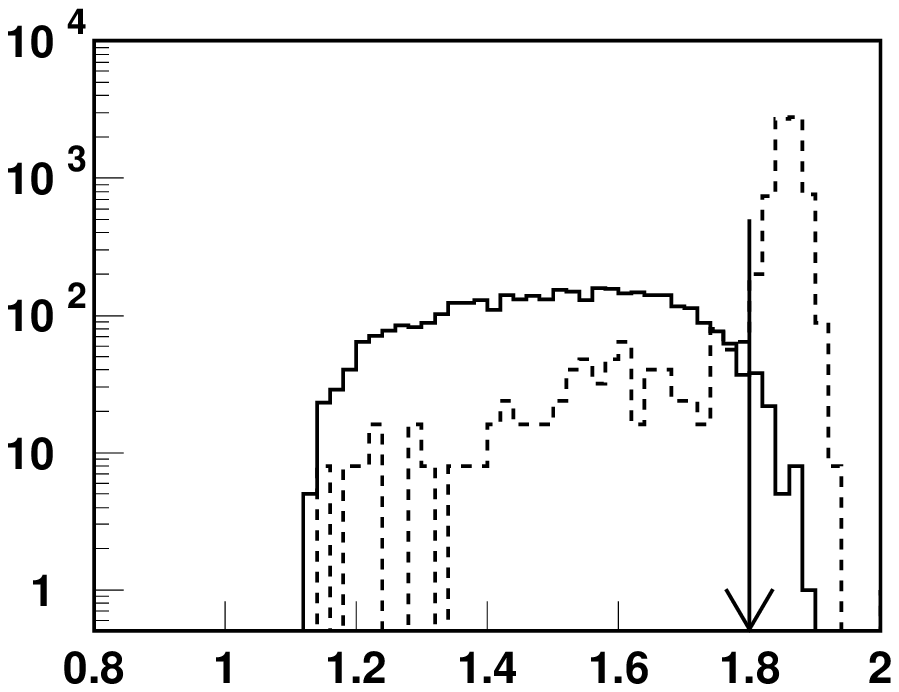}
\put(-137,13){$M_{\phi\mu^+}$ (GeV/$c^2$)}
\put(-225,45){\rotatebox{90}{Events/(0.02 GeV/$c^2$)}}
\caption{The distributions of the invariant masses of  $\phi \mu^+$ (solid)
         and $\phi \pi^+$ (dashed) for $D^+ \to
\phi \mu^+\nu_\mu$ and $D^+ \to \phi \pi^+$ from Monte Carlo events.}
\label{dp_phimunu_phipi_invm_umiss}
\end{center}
\end{figure}

In the semileptonic decays, one
neutrino is undetected. A kinematic quantity
$U_{miss}\equiv E_{miss}-p_{miss}$ is used
to obtain the information of the missing neutrino, where
$E_{miss}$ and $p_{miss}$ are the total energy
and the total momentum of all missing particles respectively.
Monte Carlo study shows that the background modes
for $D^+ \to \phi e^+(\mu^+)\nu_{e(\mu)}$ come primarily from
$D^+ \to \phi \mu^+(e^+)\nu_{\mu(e)}$,
$D^+ \to \phi \pi^+\pi^0$ and $D^+ \to \phi \pi^+$.
The requirement $|U_{miss}| < 3 \sigma_{U_{miss}}$ is imposed
to reduce these backgrounds, where $\sigma_{U_{miss}}$ is
the standard deviation of the $U_{miss}$ distribution obtained from
Monte Carlo simulation. In Fig.~\ref{dp_phimunu_phipi_invm_umiss},
the solid and dashed histograms show respectively the distributions of
the invariant masses of $\phi \mu^+$ and $\phi \pi^+$ for
$D^+ \to \phi \mu^+\nu_\mu$ and $D^+ \to \phi \pi^+$ from Monte Carlo
events.
The candidate mode $D^+ \to \phi \mu^+\nu_\mu$
has a potential hadronic
background of $D^+ \to \phi \pi^+$  due to the misidentification
of a charged pion as a muon.
These backgrounds can be suppressed by requiring the
invariant masses of the $\phi \mu^+$ combinations to be less than 1.8
GeV/$c^2$. Similarly the invariant masses of the $\phi e^+$ combinations
are also required to be less than 1.8 GeV/$c^2$ in order to remove the
$D^+ \to \phi \pi^+$ background for the $D^+ \to \phi e^+\nu_e$
decay. To suppress backgrounds from decays with neutral pion,
for example $D^+ \to \phi \pi^+\pi^0$,
the number of the isolated photons is required to
equal two if the singly tagged $D^-$ mode
contains a $\pi^0$ meson, otherwise zero.
The events with isolated photons those energies are larger than 0.1 GeV
excluded in the tagged $D^-$ are not kept.

\begin{figure}[hbtp]
\begin{center}
\includegraphics[width=7cm,height=6cm]
{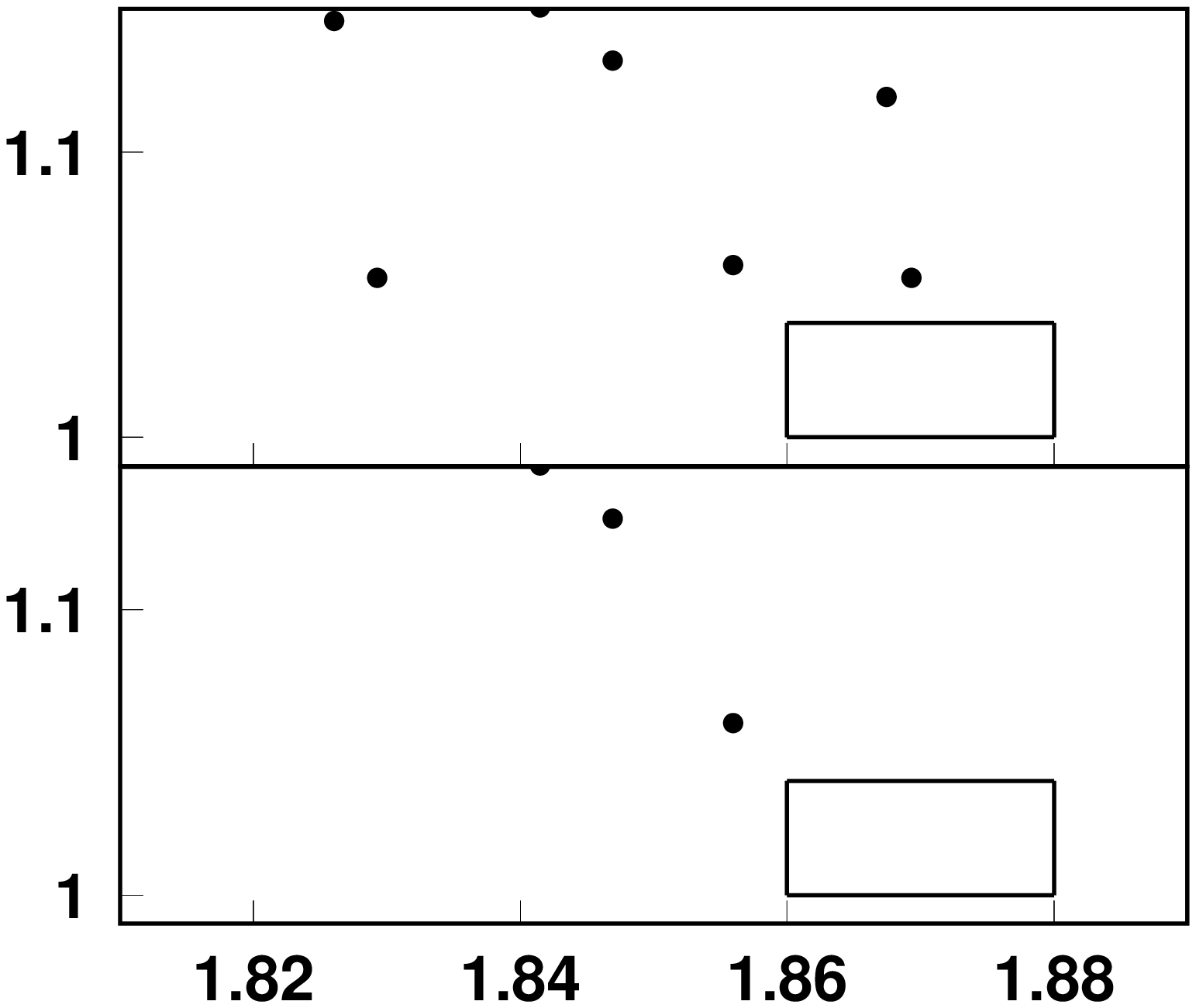}
\put(-159,130){\bf (a)}
\put(-159,65){\bf (b)}
\put(-130,-3){$M_{mKn\pi}$ (GeV/$c^2)$}
\put(-208,50){\rotatebox{90}{$M_{K^+K^-}$ (GeV/$c^2)$}}
\caption{Scatter plot of the $K^+K^-$ invariant masses
versus the $mKn\pi$ invariant masses
for the $D^+ \to \phi e^+ \nu_e$
candidates: (a) before and (b) after
applying both $U_{miss}$ and $M_{\phi e^+}$ cuts, where
the rectangle represents the combined signal region of
$\phi$ and $D^+$.}
\label{dp_phienu_2d}
\end{center}
\end{figure}

\begin{figure}[hbtp]
\begin{center}
\includegraphics[width=7cm,height=6cm]
{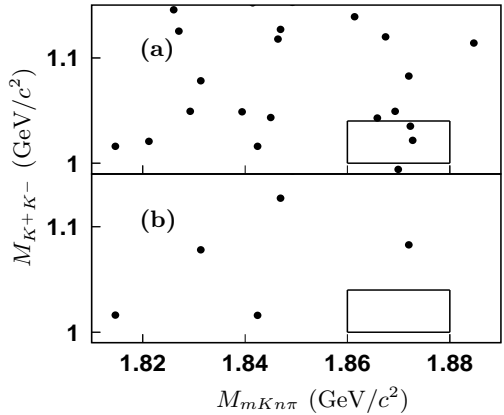}
\put(-159,130){\bf (a)}
\put(-159,65){\bf (b)}
\put(-130,-3){$M_{mKn\pi}$ (GeV/$c^2)$}
\put(-208,50){\rotatebox{90}{$M_{K^+K^-}$ (GeV/$c^2)$}}
\caption{Scatter plot of the $K^+K^-$ invariant masses versus
the $mKn\pi$ invariant masses
for the $D^+ \to \phi \mu^+ \nu_\mu$
candidates: (a) before and (b) after
applying both $U_{miss}$ and $M_{\phi \mu^+}$ cuts, where
the rectangle represents the combined signal region of $\phi$ and $D^+$.}
\label{dp_phimunu_2d}
\end{center}
\end{figure}

Fig.~\ref{dp_phienu_2d} and Fig.~\ref{dp_phimunu_2d} show respectively
the scatter plots of the invariant masses of the $K^+K^-$
combination versus the $mKn\pi$ combination
for $D^+ \to \phi e^+ \nu_e$ and $D^+ \to \phi \mu^+ \nu_\mu$
before and after
applying both $U_{miss}$ and $M_{\phi l^+}$ cuts.
No event for the $D^+ \to \phi e^+ \nu_e$ and
$D^+ \to \phi \mu^+ \nu_\mu$ decays satisfies
the selection criteria.


\subsection{Candidates for $D^+ \to \phi \pi^+$ and
$D^+ \to K^-\pi^+\pi^+$}

\begin{figure}[htbp]
\begin{center}
\includegraphics[width=8cm,height=7cm]
{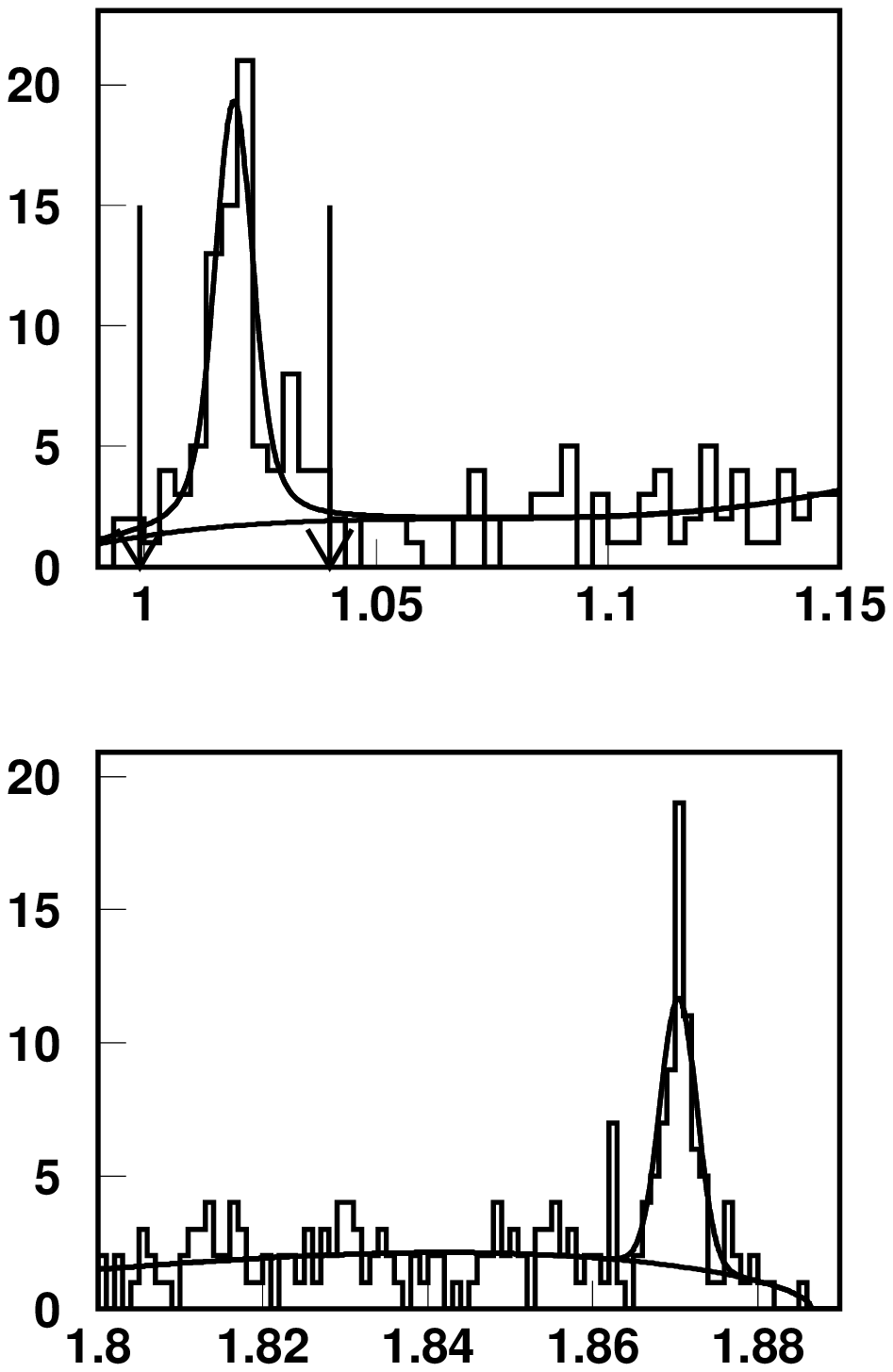}
\put(-115,157){\bf (a)}
\put(-115,70){\bf (b)}
\put(-114,95){$M_{K^+K^-}$ (GeV/$c^2)$}
\put(-114,6){$M_{\phi\pi^+}$ (GeV/$c^2)$}
\put(-226,55){\rotatebox{90}{Events/(0.001 GeV/$c^2$)}}
\caption{The distributions of the fitted invariant masses of
         (a) $K^+K^-$ combination in the $D^+$ signal region,
         (b) $K^+K^-\pi^+$ combination in the $\phi$ signal region.}
\label{dp_phipi_k2p_data}
\end{center}
\end{figure}

The candidate for $D^+ \to \phi \pi^+$  is reconstructed from
$D^+ \to  K^+K^- \pi^+$. The distribution of the fitted invariant mass
of the $K^+K^-\pi^+$ combination is shown in Fig.~\ref{dptags_9modes}(d).
As mentioned in previous subsection, the $\phi$ meson
is selected though its decay to $K^+K^-$.
Fig.~\ref{dp_phipi_k2p_data}(a)
shows the fitted invariant mass spectrum of the $K^+K^-$ pairs
from the $D^+$ signal region for the $D^+ \to K^+K^-\pi^+$ candidate
events. Fitting the mass spectrum with a Gaussian function convoluted
Breit-Wigner gives $64.3 \pm 10.3$  $\phi$ signal events.
To select the decay $D^+ \to \phi \pi^+$, we require the invariant mass
of the $K^+K^-$ combination to be within 20 MeV/$c^2$ of the nominal $\phi$
mass, and the invariant mass distribution of $K^+K^-\pi^+$ combination
is shown in Fig.~\ref{dp_phipi_k2p_data}(b), a clear signal of the decay
$D^+ \to \phi \pi^+$ is observed. Fitting
the mass spectrum of the $K^+K^-\pi^+$ combination with a Gaussian function
as the signal and a special function~\cite{psipp1,psipp2} for the background
yields $55.6 \pm 8.6$ signal events,
where the mass resolution is fixed at 0.0022 GeV/$c^2$ determined
from the Monte Carlo simulation. There may be the $K^+K^-$
combinatorial background in the observed $\phi \pi^+$ events.
The background events are estimated to be $6.1 \pm 2.5$ by
the $\phi$ sideband $|M_{K^+K^-} - 1.10| < $ 0.05 GeV/$c^2$.
For the $D^+ \to K^- \pi^+ \pi^+$ candidates, the fitted invariant mass
spectrum is shown in Fig.~\ref{dptags_9modes}(a),
and $3153.8 \pm 69.2$ events are yielded from the fit.

In the fitted yields, there are still some background contaminations
which shape a peak under the $D^+$ peak. These backgrounds
are estimated by analyzing the
Monte Carlo sample which is about 14 times larger than the data.
The Monte Carlo events are generated as $e^+e^- \to D\overline D$,
and both $D$ and $\overline D$ mesons decay to all possible
modes according to the decay modes and the branching fractions
quoted from PDG~\cite{pdg} excluding decay modes under study.
The number of events satisfying the selection criteria is then normalized
to the data. For the $D^+ \to \phi \pi^+$ decay, the dominant background
modes are $D^+ \to \phi \mu^+\nu_\mu$, $D^+ \to \phi e^+\nu_e$ and
$D^+ \to \phi \pi^+\pi^0$, and the number of the background events
is $3.6 \pm 1.1$. For the $D^+ \to K^- \pi^+\pi^+$ decay,
the number of the background events is $35.2 \pm 4.2$,
After subtracting the numbers of the background events,
$45.9 \pm 9.0$ and $3118.6 \pm 69.3$ signal events
for the $D^+ \to \phi \pi^+$ and $D^+ \to K^-\pi^+\pi^+$
decays are remained.

\section{Results}

\subsection{Monte Carlo efficiency}

The reconstruction efficiencies of the
semileptonic decays $D^+ \to \phi e^+\nu_e$, $D^+ \to \phi \mu^+\nu_\mu$
and the hadronic decays $D^+ \to \phi \pi^+$, $D^+ \to K^- \pi^+\pi^+$
are estimated by Monte Carlo simulation with the GEANT3 based
Monte Carlo simulation package~\cite{GEANT}. A detailed  Monte Carlo
study shows that the efficiencies
are $\epsilon_{\phi e^+\nu_e} = (2.46 \pm 0.03)\%$,
$\epsilon_{\phi \mu^+\nu_\mu} = (2.43 \pm 0.03)\%$,
$\epsilon_{\phi \pi^+} = (6.47 \pm 0.05)\%$ and
$\epsilon_{K^-\pi^+\pi^+} = (24.99 \pm 0.13)\%$.

\subsection{Branching fraction}

A $D^+$ signal is observed neither in the $D^+ \to \phi e^+\nu_e$
decay nor in the $D^+ \to \phi \mu^+\nu_\mu$ decay.
The upper limit of the branching fraction can be calculated using
\begin{equation}
BF < \frac{N_{sg}}{\epsilon \times N^{D^-}_{tag} \times (1 - \delta)},
\label{brlimit}
\end{equation}
where $N_{sg}$ is the upper limit of the signal yield
given with the Feldman-Cousins prescription~\cite{feldcou},
which is 2.44 for zero observed event in the absence of background
for the confidence level of 90\%, $\epsilon$ is the detection
efficiency, $N^{D^-}_{tag}$ is the number of the singly tagged
$D^-$ mesons and $\delta$ is the systematic error. The upper limits of
the branching fractions are summarized
in the second column of Table~\ref{uplbr}. Comparisons with those
obtained by MARKIII~\cite{mark} and BESI~\cite{bes1} experiments  are
also listed in the third and fourth columns of Table~\ref{uplbr}.

\begin{table}[h]
\caption{\small Upper limits for the branching fractions (\%)  are given
at the 90\% confidence level.}\label{uplbr}
\begin{center}
\begin{tabular}{cccc}
\hline\hline
Mode &~~BESII~~ & ~~MARKIII~~ & ~~BESI  \\
\hline
$D^+ \to \phi e^+\nu_e$    & $<$ 2.01 & $<$ 2.09 & 1.38\\
$D^+ \to \phi\mu^+\nu_\mu$ & $<$ 2.04 & $<$ 3.72 & -\\
\hline\hline
\end{tabular}
\end{center}
\end{table}
The systematic error, $\delta$, includes the uncertainties from
particle identification (1.4\% for $D^+\to \phi e^+\nu_e$,
1.6\% for $D^+\to \phi\mu^+\nu_\mu$),
tracking efficiency (2.0\% per track),
photon reconstruction (2.0\%), $U_{miss}$ selection (0.6\%),
the number of the singly tagged $D^-$ mesons (3.0\%) and
Monte Carlo statistics (1.2\%).
These uncertainties are added in quadrature to obtain
the total systematic errors to be 7.3\%
for $D^+ \to \phi e^+\nu_e$ and  7.4\% for
$D^+ \to \phi \mu^+\nu_\mu$.
The ratio of the branching fractions for $D^+ \to \phi \pi^+$
relative to $D^+ \to K^-\pi^+\pi^+$ can be obtained
by
\begin{equation}
\frac{BF(D^+ \to \phi \pi^+)}{BF(D^+ \to K^-\pi^+\pi^+)} =
\frac{N_{\phi \pi^+}}{N_{K^-\pi^+\pi^+}}
\times \frac{\epsilon_{K^-\pi^+\pi^+}}{\epsilon_{\phi \pi^+}},\label{one}
\end{equation}
where $N_{\phi \pi^+}$, $N_{K^-\pi^+\pi^+}$ are the numbers of the
signal events, and $\epsilon_{\phi \pi^+}$, $\epsilon_{K^-\pi^+\pi^+}$
are the detection efficiencies for the tagging and normalizing modes.
Inserting these numbers into Eq.~(\ref{one}),
the relative branching fraction is obtained
to be $0.057 \pm 0.011 \pm 0.003$, where the first error is statistical,
and the second systematic. By normalizing $D^+ \to \phi \pi^+$
to $D^+ \to K^-\pi^+\pi^+$, the systematic uncertainty
from tracking efficiency can be canceled.
The residual error, including both contributions of the tagging
and normalizing modes, arises from uncertainties in
particle identification (0.5\% for uncanceled kaon track
in $D^+ \to \phi\pi^+$, 0.5\% for uncanceled pion track
in $D^+ \to K^-\pi^+\pi^+$),the number of the singly
tagged $D^+$ meson (3.7\% for $D^+ \to \phi \pi^+$, 3.3\% for $D^+ \to
K^-\pi^+\pi^+$), background uncertainty (1.4\% for $D^+ \to \phi \pi^+$,
0.2\% for $D^+ \to K^-\pi^+\pi^+$) and Monte Carlo statistics (0.8\%  for
$D^+ \to \phi \pi^+$, 0.5\% for $D^+ \to K^-\pi^+\pi^+$).
These uncertainties are added in quadrature to obtain
the total systematic error to be 5.3\%. The relative branching
fraction is compared with other measurements
in Table~\ref{ratiowidth}.
\begin{table}[htbp]
\caption{\small Ratio of partial widths $\frac{\Gamma(D^+ \to
\phi \pi^+)}{\Gamma(D^+ \to K^-\pi^+
\pi^+)}$ and comparison with other
experiments (Evts stands for $D^+ \to
\phi \pi^+$).}\label{ratiowidth}
\begin{center}
\begin{tabular}{ccc} \hline\hline
Reference & $\frac{\Gamma(D^+ \to \phi \pi^+)}{\Gamma(D^+ \to K^-\pi^+
\pi^+)}$ & Evts\\
\hline
BESII(this work)     & $0.057 \pm 0.011 \pm 0.003$ & 46 \\
E687~\cite{E687_1}     & $0.058 \pm 0.006 \pm 0.006$ & -    \\
WA82~\cite{WA82}     & $0.062 \pm 0.017 \pm 0.006$ & 19   \\
CLEO~\cite{CLEO1992} & $0.077 \pm 0.011 \pm 0.005$ & 128  \\
NA14~\cite{NA14}     & $0.098 \pm 0.032 \pm 0.014$ & 12   \\
E691~\cite{E691_1}     & $0.071 \pm 0.008 \pm 0.007$ & 84   \\
MARKIII~\cite{MARKIII} & $0.084 \pm 0.021 \pm 0.011$ & 21   \\
\hline\hline
\end{tabular}
\end{center}
\end{table}

In addition, the branching fraction for $D^+ \to \phi \pi^+$
is obtained to be
\begin{equation}
BF(D^+ \to \phi \pi^+) = (5.2 \pm 1.0 \pm 0.4) \times 10^{-3}
\end{equation}
by using the branching fraction of the $D^+ \to K^-\pi^+\pi^+$ decay
quoted
from PDG~\cite{pdg}, where the first error is statistical and the second
systematic. The systematic error of the branching fraction for
 $D^+ \to \phi \pi^+$ includes the systematic error
of the ratio $BF(D^+ \to \phi \pi^+)/BF(D^+ \to K^-\pi^+\pi^+)$ ( 5.3\%)
and the uncertainty of the branching fraction for
$D^+ \to K^-\pi^+\pi^+$ (6.5\%). The total systematic
error is obtained to be 8.4\% by adding these uncorrelated errors
in quadrature.

\section{Summary}

Using a data sample of integrated luminosity of 33 $pb^{-1}$
collected around 3.773 GeV with the BESII detector at the BEPC,
the semileptonic decays $D^+ \to \phi e ^+\nu_e$,
$D^+ \to \phi \mu^+\nu_\mu$ and the hadronic decay
$D^+ \to \phi \pi^+$ are studied. The upper limits of the branching
fractions are set to be $BF(D^+ \to \phi e ^+\nu_e) <$ 2.01\% and
$BF(D^+ \to \phi \mu^+ \nu_\mu) <$ 2.04\% at the 90\% confidence level.
The ratio of the branching fractions for $D^+ \to \phi \pi^+$ relative
to $D^+ \to K^-\pi^+\pi^+$ is
measured to be $0.057 \pm 0.011 \pm 0.003$. In addition,
the branching fraction for $D^+ \to \phi \pi^+$ is obtained
to be $(5.2 \pm 1.0 \pm 0.4) \times 10^{-3}$.

\vspace*{1.0cm}

{\large \bf Acknowledgments}

\vspace*{0.5cm}

The BES collaboration thanks the staff of BEPC for their hard
efforts. This work is supported in part by the National Natural
Science Foundation of China under contracts Nos. 10491300,
10225524, 10225525, 10425523, the Chinese Academy of Sciences
under contract No. KJ 95T-03, the 100 Talents Program of CAS under
Contract Nos. U-11, U-24, U-25, the Knowledge Innovation Project
of CAS under Contract Nos. U-602, U-34 (IHEP), the National
Natural Science Foundation of China under Contract  No. 10225522
(Tsinghua University).

\end{document}